# Partial Differential Phase Shift Keying – Theory and Motivation

Er'el Granot[1,2] and Shalva Ben-Ezra[1]

1 : Finisar-Israel Ltd. , 3 Golda. Meir, Nes-Ziona, 74140, Israel. email: erel@ariel.ac.il
2: Department of Electrical and Electronics Engineering, Ariel University center of Samaria, Ariel, Israel

**Abstract**

Recently, many evidences demonstrate that partial Differential Phase Shift Keying (i.e., when the delay inside the Delay Interferometer is shorter than the symbol period) can partially compensate the signal deformation caused by spectrally narrowing the optical channel (by interleavers, add-drop elements, WDM filters, etc.). In this paper the source of this effect is investigated with numerical simulations and, to the best of our knowledge for the first time, *analytically*. We found that our analytical analysis matched the simulation results with high accuracy. Furthermore, a phenomenological relation, which relates the optimum Free Spectral Range to the channel bandwidth, was derived.

**Introduction.** Differential Phase Shift Keying (DPSK) has several benefits over the more ordinary On-Off-Keying (OOK) modulation format. It is less susceptible to chromatic dispersion and nonlinear effects; moreover, the main benefit of this method is detected when using a balanced receiver which yields OSNR sensitivity enhancements of up to 3dB [1].

Yet, DPSK signal detection requires an optical demodulator, usually based on an unbalanced Delay-Interferometer (DI) [2]. Until recently it was common knowledge that the best performances are achieved with a delay $\Delta T = nB^{-1}$, where $n$ is an integer number and $B$ is the symbol rate (SR).

When a DPSK signal passes through multiple spectrally narrow optical filters, its quality is deteriorated. (For the impact of other sources of signal degradation, see Refs.3-10). Recently it was demonstrated, both experimentally and numerically [11,12], that the impairments due to optical spectrum narrowing can be partially compensated by using a DI having a differential delay smaller than the symbol time slot, i.e., $\Delta T < B^{-1}$, resulting in a Free Spectral Range (FSR) of the DI to be larger than the symbol rate, i.e., FSR>*B*. Moreover, it was shown[13] that the FSR increment beyond the bit-rate generally improves the system's tolerance to chromatic dispersion.

Thus, to date, a DI with FSR>*B* seems to be the best demodulator for spectrally narrow optical lines. The problem is that the source of this effect is till not completely understood. This is quite puzzling because the agreement of the experiment with simulations demonstrates that all the sources of this effect were taken under consideration. The challenge in identifying the main cause is the complexity of the system. The system includes many components: linear and non linear, with and without noises, optical and electrical, and the relations between them are not trivial.

Aside from the fact that this problem has practical significance, it also appears as a fundamental issue. The question is not about the Bit Error Rate (BER) or the OSNR Penalty of the optical filter, the real question should be "what is the optimal FSR for a given Bandwidth (BW) filter?" The FSR and the filter's BW have the same dimensions (Hz), and therefore, by normalizing them to SR, the solution is reduced to a *dimensionless,* and therefore *generic* (in the sense that they are SR independent), curve. It is the object of this paper to investigate theoretically the effect, to explain the validity of the linear interpretation, to formulate exactly the linear approximation, and to derive a more accurate expression for the curve: optimal FSR vs. filter's BW.

**Problems with the linear approximation**: One can naively say that if the optical spectrum is narrower, then the DI should be spectrally wider (and thus the delay should be shorter). Although this view yields eventually a curve, which is similar to the accurate one (we show that at the end of the paper), it regards the entire detection system as a linear one, while it is clearly not. A DPSK has a balanced photo-detector (BPT), i.e., both exits of the DI are measured. If more spectral energy passes through one exit, then *less* passes through the second one. The linear interpretation can explain the constructive arm of the DI, but fails in the destructive one. Mathematically, a cosine filter (the constructive arm) can partially compensate a narrow filter, but a sine filter cannot.

Moreover, since the two detectors measures power (the square of the field) and are subtracted then the final detection is the *product* of the two exits. This is definitely not a linear device.

To emphasize the complex behaviour of the DI with the BPT, let us assume that we sent a sign function, i.e., $s(n) = \text{sgn}(n) = [...-1,-1,-1,-1,+1,+1,+1,+1,...]$. When this signal passes through the filter, its exit can be written as $S(t\Delta f)$ (where $\Delta f$ is the filter's bandwidth, and $S(x)$ stands for a smooth step function). Beyond the DI the electrical signal is proportional to $S[(t+\Delta T/2)\Delta f]S[(t-\Delta T/2)\Delta f]$, and since the sampling is taken at the center of the symbol ($t=0$ in this case) then the sampling measurement is a function of the product $\Delta T \Delta f$. This means that if the filter width $\Delta f$ decreases, then the DI delay $\Delta T$ should respectively *increase*, which contradicts the experimental result. From this simple example it seems that not only does the conduct of the DI deviate from linearity but in some cases it even contradicts it. How come then that the linear derivation yields eventually a reasonable curve?

**The model.** One way to proceed is to realize that this effect is independent of noise. The BER is of course noise dependent but the characteristic optimal FSR vs. filter's BW is not. This realization reduces the problem to finding the FSR, which maximizes the eye-opening as a function of the BW.

For a wide filter any increment of the DI's FSR narrows the eye-pattern and clearly deteriorates the system's performance. In Fig.1 we illustrate this point for BW=3*B*.

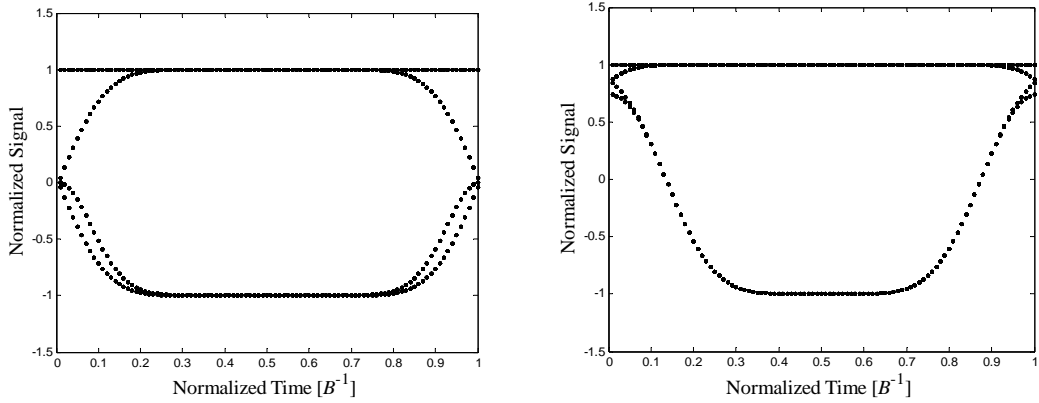

Fig.1: Eye pattern for a wide (BW=3B) filter. The left figure illustrates FSR=B, and the right figure illustrates FSR=1.35B.

However, when the BW is narrow (narrower than the BR) then we see that the eye becomes more open (Fig.2)

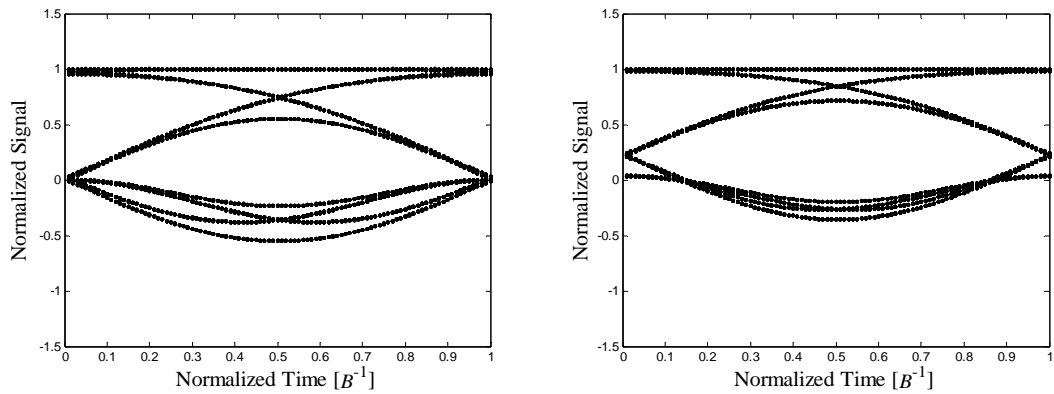

Fig.2: Eye pattern for a narrow (BW=0.6B) filter. The left figure illustrates FSR=$B$, and the right figure illustrates FSR=1.35$B$.

By examining carefully the contribution of the different structures of a random sequence to narrowing the eye we find that the eye-opening is basically determined by two structures.

**The dynamics of the minima of the eye-pattern.** The minima of the EO is determined by a section of the sequence, which oscillates between two successive symbols, i.e., $s(n) = (-1)^n = [...+1,-1,+1,-1,+1,-1,+1,-1,...]$.

After passing through the filter the signal resembles a harmonic function (see Figs.3 and 4).

For simplicity we choose a Gaussian filter:

$$H(\omega) = \exp[-(\omega/\delta)^2] = \exp\left[-\left(\frac{\omega}{\Delta f}\frac{\sqrt{\ln 2}}{\sqrt{2\pi}}\right)^2\right] \quad (1)$$

Where $\Delta f$ is the FWHM of the filter and for convenience matters we use $\delta \equiv \sqrt{2/\ln 2}\,\pi\Delta f$. We chose Gaussian filters since our analysis and the result of many simulations show that the differences between the filters have a minor impact on the final curve. Commercial filters usually have a super Gaussian shape, but it will be shown that despite the simplicity of the Gaussian filters, the final curve agree very well with experimental and simulation results.

Due to the filter we can take, with great accuracy, only the two first harmonics, which the sequence is consisted of:

$$E(t) = \frac{4}{\pi}\cos(\pi t)\exp[-(\pi/\delta)^2] - \frac{4}{3\pi}\cos(3\pi t)\exp[-(3\pi/\delta)^2] \quad (2)$$

This optical signal after passing through the DLI and then being detected by the balanced photo detectors is converted to an electrical signal:

$$I(t) = E\left(t+\frac{\tau}{2}\right)E\left(t-\frac{\tau}{2}\right) \cong$$
$$\frac{16}{\pi^2}\cos\left[\pi\left(t+\frac{\tau}{2}\right)\right]\cos\left[\pi\left(t-\frac{\tau}{2}\right)\right]\exp[-2(\pi/\delta)^2] -$$
$$\frac{16}{3\pi^2}\left[\cos\left[\pi\left(t+\frac{\tau}{2}\right)\right]\cos\left[\pi\left(3t-\frac{\tau}{2}\right)\right] + \cos\left[3\pi\left(t+\frac{\tau}{2}\right)\right]\cos\left[\pi\left(t-\frac{\tau}{2}\right)\right]\right]\exp[-10(\pi/\delta)^2]$$

(3)

Since the minima are measured at $t=1/2$ then the minima points of the eye-pattern are

$$I_{\min}(\tau,\delta) = -\frac{16}{\pi^2}\sin^2(\pi\tau/2)\exp(-2(\pi/\delta)^2) - \frac{16}{3\pi^2}[\cos(\pi\tau)-\cos(2\pi\tau)]\exp(-10(\pi/\delta)^2)$$

(4)

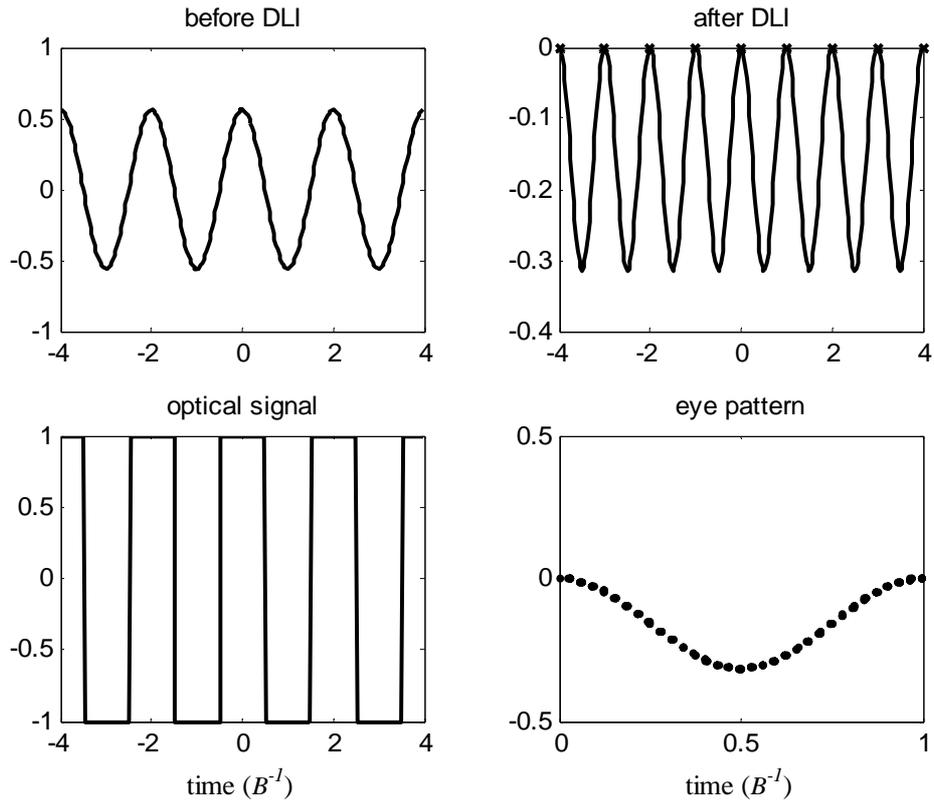

Fig.3: Detection of a train of pulses (…1,-1,1,-1,…) with an optical filter of 0.65*B* and a DLI with FSR=*B*. The original optical sequence is the sub figure on the bottom left. The sequence after the optical filter but before the DLI is the upper left figure. The upper right figure is the electrical pulse after the DLI and the balanced photodetector (the crosses stand for the decision points), and the resultant eye-pattern is at the bottom right corner.

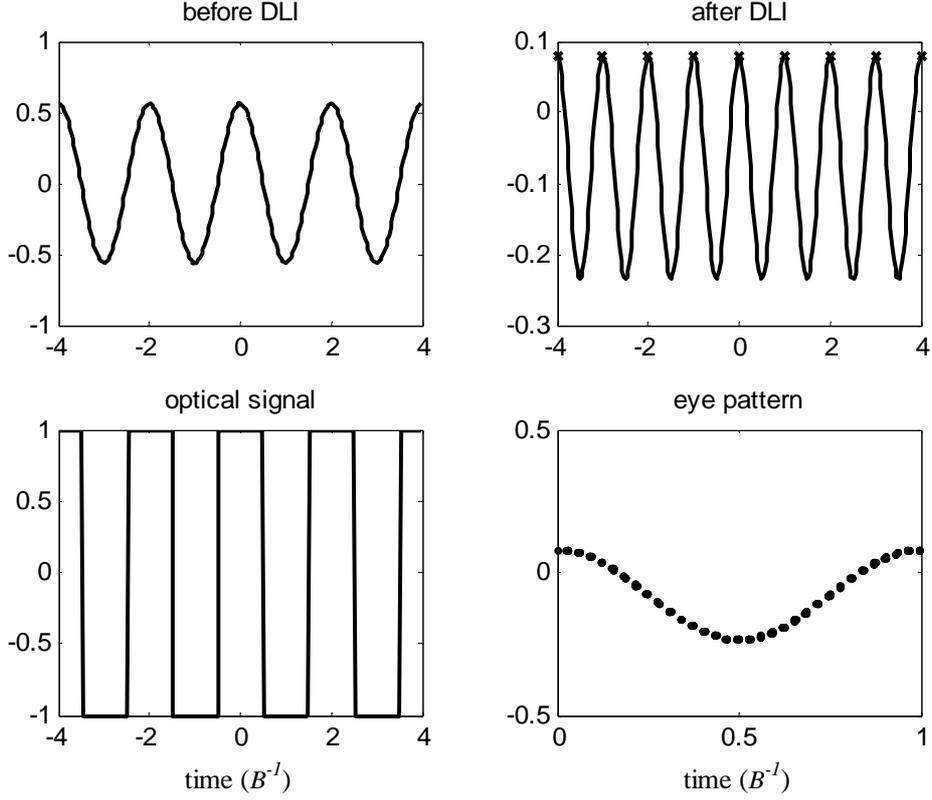

Fig.4: Same as Fig.3 but for FSR=1.5*B*.

**The dynamics of the maxima of the eye-pattern.** On the other hand, the maxima of the EO is determined by a sub-sequence, which consist of two successive identical symbols, i.e., $s(n)=[...-1,-1,-1,+1,+1,-1,-1,-1,...]$ (see Figs.5 and 6).

After passing through the filter the field can be written:

$$E(t)=\frac{2}{\pi}\int_{-\infty}^{\infty}\frac{\sin(\omega)}{(\omega/2)}\exp\left[i\omega t-\left(\frac{\omega}{\delta}\right)^2\right]d\omega-1=\mathrm{erf}\left[\left(t+\frac{1}{2}\right)\frac{\delta}{2}\right]-\mathrm{erf}\left[\left(t-\frac{1}{2}\right)\frac{\delta}{2}\right]-1 \quad (5)$$

where erf stands for the Error function.

This optical signal after passing through the DI and then being detected by the balanced photo detectors is converted to an electrical signal, which is proportional to:

$$I(t)=E\left(t+\frac{\tau}{2}\right)E\left(t-\frac{\tau}{2}\right)=$$
$$\left\{\mathrm{erf}\left[\left(t+1+\frac{\tau}{2}\right)\frac{\delta}{2}\right]-\mathrm{erf}\left[\left(t-1+\frac{\tau}{2}\right)\frac{\delta}{2}\right]-1\right\}\left\{\mathrm{erf}\left[\left(t+1-\frac{\tau}{2}\right)\frac{\delta}{2}\right]-\mathrm{erf}\left[\left(t-1-\frac{\tau}{2}\right)\frac{\delta}{2}\right]-1\right\}$$

(6)

Therefore, at the maxima point (t=0) of the eye-pattern:

$$I_{\max}(\tau,\delta) = \left\{ \operatorname{erf}\left[\left(1+\frac{\tau}{2}\right)\frac{\delta}{2}\right] - \operatorname{erf}\left[\left(\frac{\tau}{2}-1\right)\frac{\delta}{2}\right] - 1 \right\}^{2} \tag{7}$$

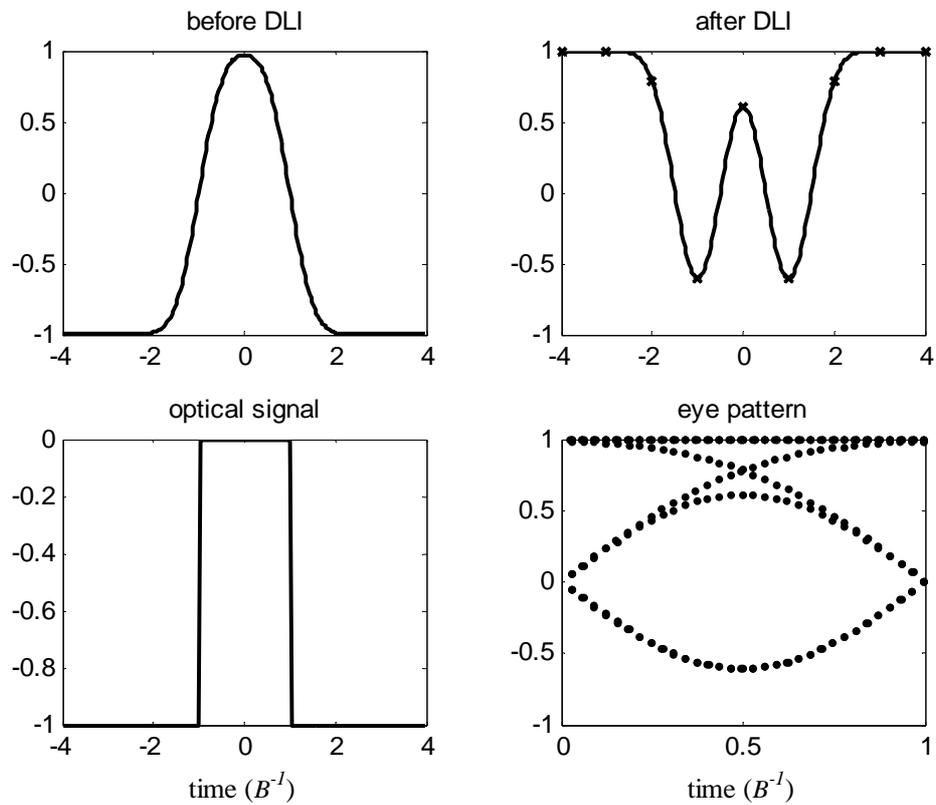

Fig.5: Detection of two successive ones (…-1,-1,-1,1,1,-1,-1,-1…) with an optical filter of 0.65*B* and a DLI with FSR=*B*. The original optical sequence is the sub figure on the bottom left. The sequence after the optical filter but before the DLI is the upper left figure. The upper right figure is the electrical pulse after the DLI and the balanced photodetector (the crosses stand for the decision points), and the resultant eye-pattern is at the bottom right corner.

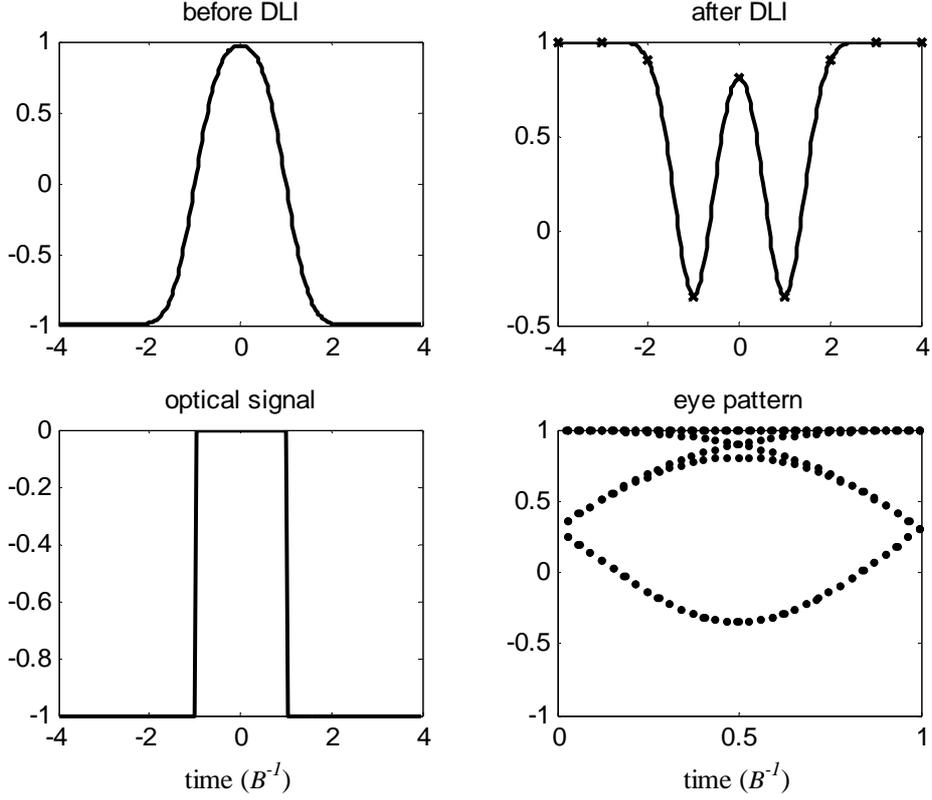

Fig.6: Same as Fig.5 but for FSR=1.5*B*.

Since in the first approximation the electrical Q is proportional to the eye opening then:

$$Q(FSR, \Delta f) \propto I_{max}\left(\frac{1}{FSR}, \sqrt{2/\ln 2}\pi\Delta f\right) - I_{min}\left(\frac{1}{FSR}, \sqrt{2/\ln 2}\pi\Delta f\right) \qquad (8)$$

The best (or optimal) FSR can be evaluated by maximizing $Q(FSR, \Delta f)$.

**Analytical solution vs. simulations.** In Fig.7 we compare the optimal FSR vs. optical filter BW of our simulation (solid curve), the experimental and simulations from the literature (circles and squares) to the analytical analysis (dashed curve). It should be noted that the fact that our curves coincide with the simulation results of Ref.[12], where a second order Gaussian was chosen, only reinforce our preconceptions that this effect is approximately insensitive of filter's shape.

Moreover, the similarity between the theory and the simulation indicates that indeed the effect is governed by a compensation between two sequences deterioration. This is a generic effect, which is independent of noise and almost insensitive of the specific filter's shape.

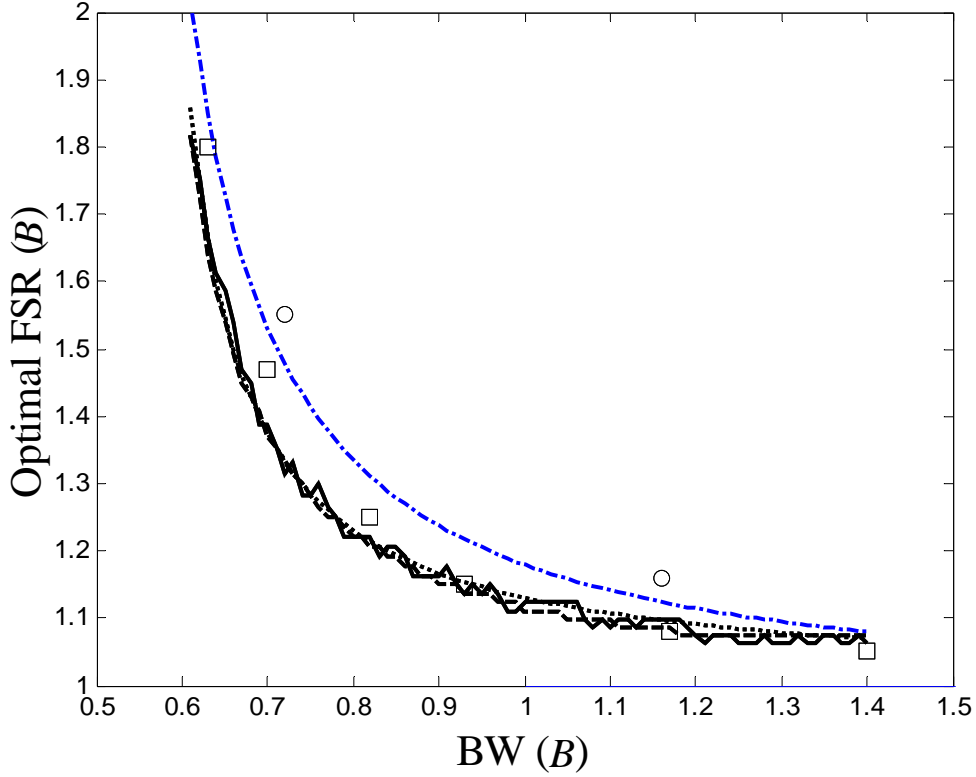

Fig.7: Comparison between different optimal curves (OFSR vs. filter's BW). The circles stand for experimental Mikkelsen et. al. [11] results, the squares stand for Malouin et. al. [12] simulations. The solid line results from the optimization of the eye-pattern (without noises); the dashed line stands for the analytical analysis, the dotted line represent formula (9) and the dash-dotted line represents the linear approximation (Eq.10).

**A phenomenological approximation and linear one.** The resultant curve allows us also to find a phenomenological approximation for the optimal FSR vs. filter's BW curve.

In Fig.7 the resultant curve of the simulation is compared to the phenomenological curve:

$$Optimal\ FSR \sim 1 + \frac{0.06}{BW - 0.54} \tag{9}$$

(in Eq.9 both the BW and the FSR are normalized to the bit-rate $B$).

Note that since the filter's BW, the FSR, and the BR are all measured in frequency units then Eq.9 is totally dimensionless.

If we regard the DI as a linear filter by taking into account only its constructive port (i.e., its destructive port is ignored), and if the optical filter is (1) and the DI is also an optical filter with a transfer function $H(\omega) \propto \cos(\omega/2FSR)$, then a straightforward analysis, shows that the DI can partially compensate the filter when choosing

$$Optimal\ FSR \sim \left[1 - 4\ln 2(BW\pi)^{-2}\right]^{-1/2} \qquad (10)$$

The two functions (9) and (10) are qualitatively similar, but the difference between them diverges for narrow bandwidth (see Fig.7). This is not surprising since, as was explained at the beginning of the paper, the DI cannot be regarded as a linear filter. Yet, since this effect is qualitatively insensitive of the filter shape, the linear approximation is a good approximation.

**Summary**. The effect, where larger FSR can improve BER for spectrally narrow channels, is investigated. It is shown that by optimizing the eye-opening of a noiseless signal an excellent estimation of the optimal FSR is achieved. We also find the exact curve of this effect by analyzing *analytically* the influence of the sub-sequences, which cause the effect. To the best of our knowledge this is the first time that this effect was addressed analytically. This analytical analysis yields an excellent match to the simulation. To complete the discussion, a simple formula was derived for the prediction of the best FSR for a given spectral BW channel.

**References**

1. A. H. Gnauck et al.and P. J. Winzer, IEEE Journal of Lightwave Technology, **23,** 115 (2005)
2. F. Seguin and F. Gonthier, "Tuneable all-fiber, delay-line interferometer for DPSK demodulation," in Proc. OFC 2005, paper OFL5, Anaheim, CA (2005).

3. H. Kim and P. Winzer, "Robustness to laser frequency offset in direct-detection DPSK and DQPSK Systems," J. Lightwave Technol. **21**, 1887-1891(2003).

4. P. Winzer and H. Kim, "Degradations in balanced DPSK receivers," IEEE Photon. Technol. Lett. **15**, 1282-1284 (2003).

5. K. P. Ho, "The effect of interferometer phase error on direct-detection DPSK and DQPSK signals," IEEE Photon. Technol. Lett. **16**, 308–310 (2004).

6. G. Bosco and P. Poggiolini, "The impact of receiver imperfections on the performance of Optical Direct- Detection DPSK," J. Lightwave Technol. **23**, 842–848 (2005).

7. J. P. Gordon and L. F. Mollenauer, "Phase noise in photonic communications systems using linear amplifiers," Opt. Lett. **15**, 1351–1353 (1990).

8. E. Iannone, F. S. Locati, F. Matera, M. Romagnoli, and M. Settembre, "High-speed DPSK coherent systems in the presence of chromatic dispersion and Kerr Effect," J. Lightwave Technol. **23**, 842–848 (2005).

9. J. Wang and J. M. Kahn, "Impact of chromatic and polarization-mode dispersions on DPSK systems using interferometric demodulation and direct detection," J. Lightwave Technol. **22**, 362–371 (2004).

10. Y. K. Lize, L. Christen, P. Saghari, S. Nuccio, A.E. Willner, R. Kashyap, and Paraschis, "Implication of Chromatic dispersion on frequency offset and Bit delay mismatch penalty in DPSK demodulation," in Proc. ECOC 2006, paper Mo3.2.5, Cannes, France (2006).

11. B. Mikkelsen, et al. Electronics Letters, **42**, 1363 (2006)

12. C. Malouin et al., IEEE Journal of Lightwave Technology, **25,** 3536 (2007)

13. Y. Keith Lizé, L. Christen, X. Wu, J-Y Yang, S. Nuccio, T. Wu, A. E. Willner, R. Kashyap, " Free spectral range optimization of return-to zero differential phase shift keyed demodulation in the presence of chromatic dispersion ", Opt. Express **15**, 6817 (2007)